\documentclass[]{aastex631}

\usepackage{amsmath} 
\usepackage{amssymb} 
\usepackage{multirow} 
\usepackage{subfigure}
\usepackage{booktabs}
\usepackage{soulutf8} 

\newif\ifShowEnglish\ShowEnglishtrue

\newcommand{\FeHlt}[1]{$\mbox{[Fe/H]}<{#1}$}

\usepackage{ulem}

\begin{document}

\title{A Catalog of 12,766 Carbon-Enhanced Metal-Poor Stars from LAMOST DR 8}

\author{Ziyu Fang}
\affiliation{School of Computer Science, South China Normal University, 510631 Guangzhou, People's Republic of China}

\author[0000-0003-3182-6959]{Xiangru Li}
\affiliation{School of Computer Science, South China Normal University, 510631 Guangzhou, People's Republic of China}

\author[0000-0002-0389-9264]{Haining Li}
\affiliation{Key Lab of Optical Astronomy, National Astronomical Observatories, Chinese Academy of Sciences, A20 Datun Road, Chaoyang, 100102 Beijing, People's Republic of China}



\begin{abstract}

Metal-poor stars are a rare and ancient type of stars; Carbon-Enhanced Metal-Poor (CEMP) stars are a subset of these celestial bodies that show an enrichment of carbon relative to iron. They are believed to be formed from gas polluted by the first generation of stars after the Big Bang and are important objects for studying the early universe, galaxy evolution, and nucleosynthesis. Due to their rarity, the search for metal-poor stars and CEMP stars is a valuable task. This study investigates the search for CEMP stars based on the low-resolution stellar spectra from LAMOST DR8, and proposes a deep learning scheme. From the LAMOST DR8 spectral library, this work discovered 12,766 CEMP star candidates. For ease of reference and use, we provide the estimated parameters $T_\texttt{eff}$, $\log~g$, [Fe/H] and [C/H] for them.

\end{abstract}





\section{Introduction} \label{sec:intro}

Stars that were born in the early universe usually tend to have lower metallicities (e.g., with \FeHlt{-2.0}
\footnote{The standard notation of [X/H]$=\log(X/H)_{\star}-\log(X/H)_{\odot}$ is adopted.}), and are referred to very metal-poor stars \citep{ARAA:Beers:2005}. They preserve important clues to the first generation of stars in the universe,
and are regarded as fossil records of the early evolution of the Galaxy.
Detailed chemical abundance patterns of these objects can be compared to
theoretical models and constrain the nucleosynthesis of early generations of supernovae \citep{Nomoto2013ARAA},
while their observed abundance trends along metallicities provide essential information about the chemical history of the Milky Way \citep{Frebel&Norris2015ARAA}.

In the past decades, significant effort has been devoted to searching for very metal-poor stars and exploring their properties through survey projects
and high-resolution follow-up spectroscopy\citep{Aoki2013AJ,Norris2013ApJ,yoon2016observational,lee2017chemical,lee2019chemical,yoon2018galactic,Venn2020MNRAS,whitten2021photometric,LiHaiNing2022ApJ,2023MNRASLucey}. One of the important discoveries about very metal-poor stars is that large proportion of these objects exhibit significant excess of carbon,
and they are referred to as Carbon-Enhanced Metal-Poor (CEMP) stars \citep{ARAA:Beers:2005,ApJ:Aoki:2007}. Moreover, the fraction of CEMP stars becomes higher when it goes to lower metallicities. The CEMP star occurence fraction could reach over 20\% at \FeHlt{-2.5} \citep{placco2014carbon},
and the high occurence makes CEMP stars the most pristine objects in the universe. Therefore, the abundance patterns and origins of CEMP stars are very important tracers to understand the early chemical evolution \citep{BonifacioAA2012}.

Based on their overabundance characteristics for neutron-capture elements, CEMP stars can be divided into different categories
such as CEMP-no, CEMP-s, CEMP-r, and CEMP-r/s stars \citep{ARAA:Beers:2005}. Although great effort has been devoted to understanding the abundance pattern of the observed CEMP stars, there are still quite many unsolved questions on their origins.
For CEMP-s stars, it is commonly believed that they originate from binary systems with an AGB companion. However, some of them were identified as isolated stars through radial velocity monitoring observations.
The formation mechanisms of the abundance pattern of neutron-capture elements in these stars remain an enigma \citep{AA:Hansen:2016,CAA:Wangshuang:202249}.
For CEMP-r/s stars, researchers have successively proposed a number of scenarios concerning their origins, including binary systems \citep{ApJ:Lucatello:2005,AA:Hansen:2016}, Dual Core-Flash Neutron Superburst theories \citep{PASA:lugaro:2009}, and enrichment through the so-called i-process \citep{AA:Abate:2016,JAA:Goswami:2020}. However, none of these models can perfectly explain the abundance patterns observed in these stars.
While for CEMP-r stars, the lack of observational data significantly limits the exploration for their origins \citep{ApJ:Safarzadeh:2019,CAA:Wangshuang:202249}. CEMP-no stars tend to dominate the lowest metallicity region, and thus their properties are very closely related to the nature of first stars and early chemical enrichment. The predominant hypotheses explaining the observed patterns in CEMP-no stars include rotating massive stars, faint supernovae, and inhomogeneous metal mixing. However, a definitive consensus remains elusive \citep{LiHaiNing2022ApJ}.

Fully understanding the nature and origin of these most ancient CEMP stars would help us explore the first stars and chemical enrichment of the early universe. However, due to their rarity and some challenges in identifying them using low-resolution spectroscopy, there are not yet much systematic searching for CEMP stars. There have been a few efforts concerning solving this problem. The first identification of CEMP stars was done by \citet{beers1992search}, from the HK low-resolution spectra, with quantitative carbon estimates provided by \citet{rossi2005estimation}. Then \cite{2010AJPlacco} selected CEMP candidates from the HES data through quantitatively estimating the strength of G-band, and later on, \cite{LiHaiNing2018ApJS} adopted similar methods to identify CEMP candidates from LAMOST DR3. Both of them have resulted in over 600 candidate CEMP stars.  Furthermore, \cite{2023MNRASLucey} compiled a CEMP star catalogue based on the slitless spectra from Gaia.
Additionally, advancements in photometric techniques, particularly with the use of narrow-band filters in J-PLUS and S-PLUS, have enhanced our ability to identify CEMP stars and estimate the [C/Fe]. For instance, \cite{whitten2021photometric} utilized S-PLUS Data Release 2 to estimate effective temperatures and metallicities for over 700,000 stars, constraining the metallicity distribution function of K-dwarfs and identifying 364 candidate CEMP stars. Following this, \cite{yang2022j} leveraged J-PLUS data to provide effective temperatures, surface gravities, and elemental abundances for approximately 2 million Galactic stars, showcasing the potential of photometric data in studying chemical evolution. More recently, \cite{huang2024j} provided effective temperature, surface gravity, age, and metallicity parameters for about 5 million stars, contributing valuable data for understanding the assembly and chemical evolution of the Milky Way. However, these studies did not quantitatively estimated the C abundance for these stars. To this end, this work is to find some CEMP star candidates by estimating C abundance based on LAMOST observations.

The Large Sky Area Multi-Object Fiber Spectroscopic Telescope (LAMOST, also called the Guo Shou Jing Telescope) is a reflecting Schmidt telescope with an effective aperture of 3.6m–4.9m  and a field of view (FOV)  5$^{\circ}$\citep{cui2012,2012_LEGUE,Zhao_2012,liu2015,2015_LAMOSTdr1}.  The focal surface of LAMOST can accommodate 4000 optical fibers and observe up to 4000 objects simultaneously.  After a two-year commissioning phase, one-year pilot survey and eleven-year regular surveys, LAMOST has collected 11,581,542 star spectra with low resolution\footnote{https://www.lamost.org/lmusers/}.
In addition, the wavelength coverage of 3700-9100{\AA} and resolution of $R\sim 1800$ enable quite robust estimation of stellar parameters for FGK stars \citep{luo2015first,Yuan2015MNRAS}, even down to very low metallicity region \citep{LiHaiNing2018ApJS,Hou2024MNRAS}.
This huge spectral database thus provides an excellent opportunity for searching for CEMP stars.

The organization of the remaining parts of this paper is as follows. Section \ref{sec:RelatedWorks_Motivation} briefly reviews the related works and discusses the motivation of this work. Section \ref{Sec:Workflow_Data_Preprocessing} introduces the overall process of CEMP star search in this paper, the reference data used, and the data preprocessing procedures. Section \ref{Sec:ParaEst_MPsearching} introduces the proposed method for estimating stellar spectral parameters. These parameter estimation results are used for screening and judging CEMP star candidates. Section \ref{Sec:CEMPrecognition} presents the CEMP star candidate search results; Section \ref{Sec:ParameterEvaluation} compares the consistency of the parameter estimation results of this work with GALAH; Section \ref{Sec:Conclusion} provides a brief summary for the proposed CEMP star search scheme and its results.

\section{Related Works and motivation}\label{sec:RelatedWorks_Motivation}

The searching technologies for CEMP stars can be divided into the following three categories: line index methods, template matching methods and machine learning methods. The afore-mentioned classification criteria is established based on their style of using the spectral information. The line index methods are characterized by the scalar description of one or more selected wavelength bands for spectral information , for example, the full width at half maximum (FWHM)  of a spectral line, or some measures based on the spectral flux integral. The template matching method is to estimate parameter or recognize objects by comparing the corresponding flux of an observed spectrum on one or several specified wavelength bands with that of some reference spectra. The machine learning method is to treat a spectrum as a vector or matrix and establish a mapping from the spectrum to the parameter or type labels. These difference between these method result in their applicability and characteristics. The line index method is the most interpretable due to the mandatory requirements for spectral lines and wavelength band selection, but it also requires the highest spectral resolution; the template matching method is sensitive to spectral quality due to the direct dependence on individual flux.  In essence, the key feature of machine learning methods is the automatic extraction of spectral features through algorithms; the automatic feature extraction can exploit the correlation between various wavelength bands and suppress some negative effects from noise. In theory, therefore, machine learning method can deal with the spectra with  relatively low quality and resolution, but need a large amount of reference data as a carrier of empirical knowledge.

 Therefore, proposal, development, and selection of these schemes depend not only on the physical interpretability of the models, the signal to noise ratio and resolution of the data, the accumulation of CEMP star observation data, but also on the available computing resources. The following part of this section reviews the studies based on these three types of methods. Although this work deals with low-resolution spectra, this review is dedicated to a comprehensive summary of CEMP searching techniques. Therefore, it is not limited to the research works based on low-resolution spectroscopy.

\subsection{Line Index Methods}\label{sec:Relatedworks:IndexMethods}

CEMP stars are typically characterized by their low metallicity [Fe/H] and high carbon abundance [C/Fe]. The line index method is based on the detection and description of spectral lines sensitive to metallicity and carbon abundance. The vast majority of early identified CEMP stars were discovered from the metal-poor star candidates in the OBJECTIVE-PRISM SURVEY, for example, the HK survey \citep{1985AJBeers,1992AJBeers} and the Hamburg/ESO survey (HES) \citep{2006ApJFrebel,2008AAChristlieb}. Both surveys search objects based on the presence of weak Ca II lines. \citet{2001AAChristlieb} published a HES catalog of some stars with strong carbon molecular lines, and subsequent medium-resolution spectroscopic observations and inspection of these objects show that at least 50\% of the stars have the characteristics of CEMP stars \citep{2006MNRASGoswami}.  However, these carbon-rich candidates are sifted based on the aggregation evaluation of a series of carbon molecular lines, e.g., CN, C2, and CH, etc. Therefore, this approach overemphasizes the cooler stars. Unfortunately, the CEMP stars with effective temperatures above 5500K usually exhibit only the unusual strength on a single carbon signature, the CH G-band near 4300\AA. Therefore, this sifting strategy likely results in the omission of such CEMP stars.

As a molecular absorption feature of C2, the Swan bands are  spectral signatures of carbon over-abundance, and can be used to identify CEMP stars. Therefore, \cite{2018MNRASCotar} proposed a supervised method and an unsupervised detection method for the Swan bands.  The supervised detection method enables the detection of Swan band features near 4737\AA ~from a high-resolution spectrum, and description of their shapes and intensity. Using the intensity of  the perceivable Swan band features, CH stars can be detected. \cite{2018MNRASCotar} then sifted the CEMP candidates based on visual detection and unsupervised machine learning methods (more about the unsupervised detection method are explored in section \ref{sec:Relatedworks:machinelearning}).

Therefore, the searching technologies based on CH G-band for CEMP stars is an important exploration direction. \citet{1999AJBeers} defined a line index, GP, on wavelength range of 4297.5\AA-–4312.5\AA  ~for the CH G-band. However, some studies show that the line index GP does not fully capture all of the characteristics of the carbon absorption characteristics on the CH G-band. To this end, \citet{2008AAChristlieb} defined a line index, GPHES, using the wavelength range of 4281.0\AA--4307.0\AA; \citet{2010AJPlacco} proposed an extended line index GPE (GPHES Extended) based on wavelength band 4200.0\AA --4400.0\AA, and accordingly computed a CEMP candidate catalog consisting of 669 previously unconfirmed CEMP candidates. The line index GPE describes the contrast between the CH G-band and the corresponding fitted continuum. \citet{Placco2011AJ} further improved the GPE and proposed the line index EGP. Based on the line indexes G1 and EGP, \citet{LiHaiNing2018ApJS}  further proposed a criterion, G1$>$4.0\AA ~and EGP$>$-0.7mag, for searching for CEMP star candidates from the stellar spectra with 4000K$< T_\texttt{eff}<$7000K and found 636 CEMP star candidates from the LAMOST DR3 library.  In the verification stage, the CH G-band feature near 4300\AA ~is selected.

\subsection{ Template Matching Methods}\label{sec:Relatedworks:templatematching}

The template matching method, also known as $\chi^2$ minimization, is widely used in astronomical observation signal identification, and parameter extraction. Its principle is to calculate the cumulative sum of the difference between the corresponding fluxes of the observed data and some templates/empirical data as a measure of the dissimilarity between them, and then predict the observed data according to the type or parameter information of the template which can minimize the cumulative sum. For example, this method can be used to estimate both [Fe/H] and [C/Fe], and also to identify CEMP stars. The idea of template matching method is to make prediction based on the most consistent template or empirical sample with the observation data. It is simple and intuitive. Therefore, the template matching method is widely adopted, and has become a benchmark method.

The keys to the template matching method are the construction of the templates/empirical database and the selection of the wavelength bands. The former determines which empirical samples are used to predict the observed data, while the latter determines what characteristics are used to judge the similarity between the observation and empirical data. For example, a rest wavelength range of 4000\AA--4650\AA is used to estimate [C/Fe] from SDSS data in \citep{2013LeeAJ}, this wavelength band contains the CH G-band feature. To estimate [Fe/H], search for metal-poor star candidates, and CEMP star candidates from LAMOST DR1 data, \citet{2015LiApJ} proposed to use the 4500\AA--5500\AA ~band. \citet{2017AAAguado} proposed a two-step method for sifting CEMP star candidates from BOSS, SEGUE and LAMOST spectra. This scheme is to first find some CEMP candidates by estimating the [Fe/H] using the CaII resonance line, and then remove some outliers (false candidates) and obvious erroneous results using a template matching method based on the local bands around the Balmer line. To search for extremely-poor metal stars and CEMP stars from LAMOST DR3, \citet{LiHaiNing2018ApJS} used wavelength range of 4360\AA--5500\AA.

The above-mentioned works indicate two classes of schemes for searching for CEMP star candidates: (1) first screen some carbon-enhanced stars, then identify some CEMP star candidates based on their metallicity; (2) first detect some metal-poor star candidates from a stellar spectrum library, and then identify some CEMP star candidates based on their [C/Fe] characteristics. Therefore, the finding of CEMP stars in some above-mentioned works are the by-products of searching for metal-poor stars, for example, \citet{2015LiApJ} and \citet{2017AAAguado}. The wavelength range selection depends on the selected search idea and the parameter configurations of our interested CEMP stars. The selection of line index schemes share similar theories.

\subsection{Machine Learning Investigations for CEMP Candidate Searching}\label{sec:Relatedworks:machinelearning}

The fundamental idea of machine learning methods is to automatically discover and extract the discriminative features of CEMP stars from some empirical data through some algorithms. Although the researches of machine learning-based CEMP candidate searching have attracted more and more attention with the rapid increase of spectral data volume, it is still in its infancy. Related studies can be divided into unsupervised method \citep{2017ApJSCarbon,2018MNRASCotar},  regression method \citep{2021ApJXie}, and supervised classification method \citep{2023MNRASLucey}.

The principle of the unsupervised CEMP star search method is to automatically find the aggregation relationship in some observed spectra through an algorithm. The spectra in a cluster are numerically very similar with each other, and there is a high probability that the corresponding objects come from a certain type of non-CEMP stars, or a subtype of CEMP stars. Therefore, a spectrum is likely with a high probability of CEMP star observation if it is in a cluster with a confirmed CEMP sample. For example, \citet{2018MNRASCotar} explored the search for CEMP stars from the GALAH (GALactic Archaeology with Hermes) observation spectral library based on a wavelength band 4720\AA--4890\AA ~and a t-SNE method. \citet{2017ApJSCarbon} searched for EMP (Extremely Metal-Poor), CEMP and Cataclysmic Variable stars from SDSS data based on a linked scatter plot method. This approach helps to find the spectra of some novel subtypes of CEMP candidates.

The idea of the regression method is to formulate the CEMP search problem as a parameter extraction problem. \citet{2021ApJXie} made a very good pilot exploration in this area. That work represented a spectrum with a 64$\times$64 matrix through a s-shaped folding technique, and estimated the parameters [Fe/H] and [C/Fe] from a stellar spectrum by inputing the matrix/image into a convolutional neural network. Furthermore, Very Metal-Poor Poor (VMP) and CEMP star candidates were recognized from LAMOST DR7 data based on a criteria [Fe/H] $<$ -2 and [C/Fe]$>$ 1. The model successfully found 260 of 414 known CEMP stars with a recall 62.80\%. Although there is a lot of room for improvement on the recall rate, the feasibility of such schemes is preliminarily explored. Compared to the CEMP star search works \citep{2015LiApJ,LiHaiNing2018ApJS} from LAMOST observations, the advantage of this scheme is that it can give [C/Fe] estimates from the low-resolution spectra. These estimated parameters help to select appropriate sources for high-resolution follow-up observations.

The supervised classification method characterizes the search problem for CEMP stars as a discrimination problem between a CEMP star and a non-CEMP star. The basic idea of this method is to project the observation spectrum into a certain space through an algorithm; the samples of CEMP stars and non-CEMP stars are in different regions of this space. Moreover, the mapping relationship from an observed spectrum to this space can be learned from some known CEMP star spectra and non-CEMP star spectra. Therefore, an observed spectrum can be first projected into this learned space and discriminated based on the region where they are located. For example, \citet{2023MNRASLucey} explored the CEMP candidate search problem  based on XGBoost method and Gaia spectra, and obtained a recall rate of approximately 73\%. The disadvantage of this work is the failure to give estimates of the [C/Fe] and other parameters for the CEMP star candidates from low-resolution spectra. However, the typical characteristic of this study is that it verified the feasibility of machine learning CEMP search method for the spectrum with a resolution as low as R=50. Therefore, the results of this work provide a feasibility basis for CEMP searching  from LAMOST low-resolution observation spectra with a resolution R=1800 using a machine learning method.

\subsection{Motivation}

With the development of large-scale spectroscopic surveys of SDSS, LAMOST, and Gaia, low-resolution spectroscopic CEMP star candidate search has attracted much attention. Due to the limitations of the technology used, however, these survey data have not been fully exploited.

With the decrease of spectral resolution and signal-to-noise ratio (SNR), the identifiability, detection, and description reliability of spectral lines are greatly reduced. Furthermore, in the large-scale low-resolution spectral survey, the SNR range of the spectrum is relatively wide.  For example, there are over 5.4 million spectra with SNR$_g \leq$20 in LAMOST DR9. Therefore, LASP, the official stellar spectral processing pipeline of LAMOST, uses the template matching method rather than the line index method. However, the template matching method is implemented by calculating the cumulative difference between the corresponding flux of the template data and an observation as a spectral dissimilarity measure.  This metric fails to exploit the local morphological and global morphological features in a spectrum. The lack of morphological feature extraction results that the noise interferences on individual pixels can seriously distort the final dissimilarity measure  even if the observed spectra are very consistent with the template spectra on the whole. Therefore, the application of the template matching method is severely limited  on low-SNR spectrum.  For example, although LAMOST DR9 released 11.22 million low-resolution spectra, the LASP estimated stellar parameters only for 6.18 million spectra. Moreover, LASP only provides effective temperature, surface gravity, and [Fe/H], but does not estimate  [C/Fe] or [C/H] \citep{WuYue2011RAA, luo2015first}.

Fortunately, \cite{ApJ:Lihaining:2015, LiHaiNing2018ApJS} made a very good pilot exploration and preliminarily showed the potential of searching for CEMP candidates from LAMOST low-resolution spectra. However, the afore-mentioned explorations were conducted based on the line index methods and the template matching method by limiting the analyzed spectral SNR to SNR$_g$\textgreater10 and SNR$_r$\textgreater15, and didn't estimate [C/Fe] or [C/H] from low-resolution spectra. In the low-resolution spectral library of LAMOST DR9 v1.1, up to 3.58 million stellar spectra do not meet the above-mentioned SNR requirements.

Therefore, it is necessary to explore CEMP star search techniques more suitable for low-resolution, low-SNR spectroscopy to fully exploit the scientific value of large-scale low-resolution spectral libraries. Fortunately, this issue has been explored based on machine learning schemes, such as unsupervised clustering \citep{2017ApJSCarbon,2018MNRASCotar}, supervised classification \citep{2023MNRASLucey}, and supervised regression estimation \citep{2021ApJXie}. In terms of the resolution and wavelength coverage of the analyzed spectra, \citet{2017ApJSCarbon} and \citet{2021ApJXie} provide excellent feasibility supports for exploring CEMP candidate search techniques from LAMOST low-resolution spectra based on machine learning schemes.

However, the afore-mentioned CEMP candidate search studies based on unsupervised clustering and supervised classification do not give parameter estimates for the candidates. The lack of spectral parameters have negative effects on selecting objects with high scientific value for the following high-resolution observation.  At the same time, it is shown that the available search method based on unsupervised clustering tends to miss rare CEMP candidates \citep{2018MNRASCotar}; the CEMP candidate search scheme based on supervised machine learning method in \citet{2023MNRASLucey}  has a recall ratio of only 73\%, which means a high missing percentage.  Although \citet{2021ApJXie} gave some preliminary results for CEMP candidate search and parameter estimation from LAMOST spectroscopy using machine learning methods, their recall is 260/414$ \approx$62.80\%. These results indicate that there exists  a lot of room for improving parameter estimation accuracy and CEMP candidate identification performance.

To this end, this work is to explore a suitable machine learning method for CEMP candidate search from low-resolution spectra, espectially the LAMOST observations.

\section{ Methodology, Data, and Preprocessing}

\label{Sec:Workflow_Data_Preprocessing}
\subsection{Overall Process}\label{Sec:Workflow_Data_Preprocessing:Workflow}
This work is to search for CEMP candidates from the LAMOST DR8 low-resolution spectroscopic database. For a given LAMOST stellar spectrum, we first 1) estimate its effective temperature $T_\texttt{eff}$, surface gravity $\log~g$, metallicity [Fe/H], and carbon abundance [C/H]; then, 2) select the metal-poor star candidates based on the criterion [Fe/H] $<$ -1; and lastly 3) identify some CEMP  star candidates from the metal-poor star candidates using the proposed criteria in \citet{ApJ:Aoki:2007}.
\begin{equation}\label{eq:CEMP_define}
\left\{
\begin{array}{lc}
\log \left(L / L_{\odot}\right) \leq 2.3 \land {[C / F e] \geq+0.7} \\
\log \left(L / L_{\odot}\right)>2.3 \land {[C / F e] \geq+3.0-\log \left(L / L_{\odot}\right)}
\end{array}
\right.
\end{equation}
where
 \begin{align*} L/L_\odot & \propto (R/R_\odot)^2(T_\texttt{eff}/T_{\texttt{eff}\odot})^4\\
                         & \propto (M/M_\odot) (g/g_\odot)^{-1} (T_\texttt{eff}/T_{\texttt{eff}\odot})^4,
 \end{align*}
$M_\odot$, $g_\odot$, and $T_{\texttt{eff}\odot}$ represent the mass, surface gravity, and effective temperature of the Sun, respectively.

\subsection{Reference Dataset for CEMP Star Search}\label{Sec:Workflow_Data_Preprocessing:Data4MP}
\label{parameter_data}

The parameter estimation procedure of the proposed scheme is implemented using a machine learning method. This method needs a reference set for estimating the model parameters.
According to the definition of CEMP stars in \citet{ApJ:Aoki:2007} (Section \ref{Sec:Workflow_Data_Preprocessing:Workflow}), the distinction between CEMP stars and carbon-normal metal-poor stars is described based on effective temperature, metallicity, and carbon abundance; there should be some significant differences on the spectral features between giants and dwarfs. This definition is designed based on the influence of stellar evolution on carbon abundance.
Therefore, this reference set should not only consist of a series of stellar spectra, but also provide reference information for parameters such as $T_\texttt{eff}$, $\log~g$, [Fe/H] and [C/H] for each spectrum.

For this purpose, we establish a reference set by cross-matching the LAMOST DR8 low-resolution stellar spectroscopic database with the APOGEE DR17 star catalog \citep{ApJS:Abdurro:2022}, the LAMOST-Subaru star catalog \citep{Aoki2022ApJ, LiHaiNing2022ApJ}, and the SAGA database \citep{PASJ:Suda:2008,MNRAS:Suda:2011,MNRAS:Suda:2013,PASJ:Suda:2017} with the TOPCAT \citep{taylor2005topcat} (the Tool for OPerations on catalogs and Tables).
In TOPCAT, we limit the maximum angular separation (Max Error) between two stellar position coordinates (Ra, Dec) to 3 arcseconds, and set the Match Selection so that each LAMOST spectrum has the best matching parameters in the APOGEE DR17 star catalog.
The LAMOST-Subaru catalog and the SAGA catalog complement APOGEE DR17 in terms of low metal abundance sources and CEMP stars. Each reference sample consists of one observed spectrum, and its parameter information such as $T_\texttt{eff}$, $\log~g$, [Fe/H], and [C/H]; the spectral data are from LAMOST DR8, and the parameter information is from the APOGEE DR17 catalogue, the LAMOST-Subaru catalogue, or the SAGA catalogue.

In conventional studies estimating stellar atmospheric physical parameters, researchers typically focus on the spectra in some parameteric regions frequently being sampled. However, these spectra are usually not from metal-poor stars or CEMP stars.
Reference samples of metal-poor stars are relatively rare, and their number is far less than that of non-metal-poor star samples.
Consequently, the initial reference data exhibits a significant imbalance between metal-poor stars and non-metal-poor stars.
This imbalance can lead to difficulties for the search model in perceiving the spectral features of metal-poor stars, including CEMP stars.
To address this issue, the work performed a downsampling on the spectra of non-metal-poor stars obtained from the matching.
The final reference set consists of 9755 spectra along with their corresponding parameter information.
In this reference set, there are 4723 samples of non-metal-poor stars, and 5032 samples of metal-poor stars. In the metal-poor star samples, there are  167 observations of CEMP stars.
In the 4723 non-metal-poor star samples, 4713 samples were randomly selected from common observations between LAMOST DR8 and APOGEE DR17 catalogs. Additionally, 9 samples are the common observations between LAMOST DR8 and the SAGA catalogs, while 1 sample is the common observation between LAMOST DR8 and the LAMOST-Subaru catalogs.
In the 5061 metal-poor star samples, 4433 are the common observations between LAMOST DR8 and APOGEE DR17 catalogs. Additionally, 486 samples are the common observations between the LAMOST DR8 and the LAMOST-Subaru catalogs, while 142 samples are the common observations between LAMOST DR8 and the SAGA catalogs.
The parameter ranges of this reference set are $[3615.00, 6772.54]$ K for $T_{\texttt{eff}}$, $[-0.15, 5.06]$ for $\log~g$, $[-4.38, 0.59]$ for [Fe/H], and $[-4.50, 0.90]$ for [C/H].

In the obtained reference data, CEMP stars are significantly less than carbon-normal metal-poor stars and non-metal-poor stars. To enhance the sensitivity of the parameter estimation model to the spectral features of CEMP stars, particularly those with extremely low metallicity, this study has assigned a label CEMP, non-metal-poor, or carbon-normal metal-poor to each reference spectrum (more about this can be found in section \ref{Sec:model}). These labels are represented by the numerical values -1, 0, and 1, respectively. The categorization labels for carbon-normal metal-poor and CEMP are calculated from the spectroscopic parameter information using the aforementioned criteria. The reference set consists of 167 CEMP samples, 4865 carbon-normal metal-poor samples, and 4723 non-metal-poor star samples.

The aforementioned reference set is randomly partitioned into a training set and a testing set at approximately an ratio 8:2. These two sets are respectively denoted as $S^{stars}_{tr}$ and $S^{stars}_{te}$. The sample sizes of these two datasets are 7804 and 1951, respectively.

\subsection{Data Preprocessing} \label{Sec:Workflow_Data_Preprocessing:Preprocessing}

The observational process is susceptible to various influences such as scattering, reflection, temperature variations, instrument anomalies, and other factors. These factors can introduce deviations between observed spectra and theoretical spectra. These deviations can reduce the accuracy of parameter estimation and the search for metal-poor stars and CEMP stars. Therefore, a series of preprocessing steps need to be conducted to minimize the adverse effects from these factors as much as possible.

The preprocessing steps used in this study are as follows:
\begin{itemize}
    \item The observed wavelengths of the spectra are transformed into the rest wavelength coordinate system based on the radial velocities estimated by the LAMOST pipeline.
    \item According to the common wavelength range in the rest frame, each observed spectrum is truncated. The truncated spectra are then divided into the blue range of [3800, 5700] \AA \ and the red range of [5900, 8800] \AA \ to eliminate stitching areas. Each spectrum is resampled using a linear interpolation method with a step size of 1 \AA \ for the blue end and 1.5 \AA \ for the red end. The flux sequences of the spectra in the blue and red ends are denoted as $\boldsymbol{f}_b$ and $\boldsymbol{f}_r$, respectively.
    \item To accurately fit the pseudo-continuum of the spectra, we first process the spectra using a median filter with a window size of 3 to eliminate some noise. Subsequently, we estimate the pseudo-continua respectively for the blue end and red end using the Asymmetric Least Squares (IAsLS) method \citep{2014Baseline}. These fitted pseudo-continua are denoted as $\boldsymbol f_b^c$ and $\boldsymbol f_r^c$, respectively.
    \item Finally, we obtain the normalized spectra, $\tilde{\boldsymbol f}_b(i) = {\boldsymbol f}_b(i)/{\boldsymbol f}_b^c(i)$ and $\tilde{\boldsymbol f}_r(i) = {\boldsymbol f}_r(i)/{\boldsymbol f}_r^c(i)$ by dividing each flux of the spectra by the corresponding flux of the fitted pseudo-continuum.
\end{itemize}

\section{Stellar Spectroscopic Parameter Estimation}\label{Sec:ParaEst_MPsearching}
In the CEMP star search workflow of this work (Section \ref{Sec:Workflow_Data_Preprocessing:Workflow}), the key procedure is to estimate the stellar spectroscopic parameters $T_\texttt{eff}$, $\log~g$, [Fe/H], and [C/H]. Therefore, this section first introduce the parameter estimation scheme (Table \ref{table:ParameterEstimationScheme}).

\subsection{
Multi-Scale Feature Learning and Information Exploitation across Wavelength Space}\label{Sec:model}
The key characteristics of this approach are its ability to learn multi-scale features and exploit cross-wavelength information. Therefore, it is referred to as a Multi-Scale Feature Learning and Information Exploitation cross Wavelength Network (MSFL-IECW Net) period.
In the output end of this model, there are two branches: one parameter estimation branch and one classification branch. According to the proposed workflow for CEMP star search (Section \ref{Sec:Workflow_Data_Preprocessing:Workflow}), theoretically, the MSFL-IECW Net does not need a classification branch.
However, in the reference data, CEMP stars are significantly less than carbon-normal metal-poor stars and non-metal-poor stars. Experimental studies indicate that the addition of a classification branch can enhance the sensitivity of the model to the spectral features of the samples with particularly low metallicity during the computation steps 1)-4) (Table\ref{table:ParameterEstimationScheme}), and improve the reliability of the model's parameter estimation for such samples.

\begin{table*}[ht]
\centering
\caption{
Parameter estimation model: multi-scale feature learning and cross-band information mining network.
}
\label{table:ParameterEstimationScheme}
{
\begin{tabular}{|c|c|c|}
\hline
  & \multicolumn{2}{c|}{A spectrum}  \\
\hline
\multirow{4}{*}{1)} & \multicolumn{2}{c|}{\textbf{
Multi-scale feature extraction}: Consecutively perform four layers of convolution followed by average}  \\
 & \multicolumn{2}{c|}{pooling with a scale of 2; each convolutional layer is computed based on $4N_i$ convolution kernels, }\\

 & \multicolumn{2}{c|}{where the numbers of kernels with scales 1, 3, 5, and 7 are all $N_i$,  and $i$ representing the index }\\
  & \multicolumn{2}{c|}{of the convolutional layer, ranging from 1 to 4. Here, $N_1 = 20$, $N_2 = 40$, $N_3 = 80$, and $N_4 = 160$.}\\
\hline

\multirow{2}{*}{2)} & \multicolumn{2}{c|}{\textbf{ Multi-scale feature fusion}{ : Based on 100 convolution kernels with a scale of 12, }}\\

 & \multicolumn{2}{c|}{feature fusion is performed with a convolution operation using a stride of 12.}\\
\hline

\multirow{4}{*}{3)} & \multicolumn{2}{c|}{\textbf{Information exploitation across wavelength}{: Segment the spectral information into 20 subbands with }} \\
& \multicolumn{2}{c|}{equal intervals according to wavelength, and utilize two LSTM learning modules to explore inter-band} \\

& \multicolumn{2}{c|}{
information. These two LSTM modules respectively extract low-band spectral features to complement} \\

& \multicolumn{2}{c|}{ high-band spectral features and high-band spectral features to complement low-band spectral features.} \\

\hline
\multirow{3}{*}{4)} & \multicolumn{2}{c|}{Fuse spectral features using a two-layer fully connected networks respectively consisting of } \\
& \multicolumn{2}{c|}{1025 and 512 neurons. Each neuron adopts the ReLU activation function,} \\
& \multicolumn{2}{c|}{and the batch normalization is applied after each fully connected layer computation.} \\
\hline
\multirow{6}{*}{5)} &A two-layer fully connected network respectively&
A two-layer fully connected network respectively \\
&consist of 256 neurons and 4 neurons.& consist of 256 neurons and 3 neurons.\\

 &The first layer incorporates batch normalization, &The first layer incorporates batch normalization,\\
  &ReLU activation function, and dropout &ReLU activation function, and dropout\\

  & regularization, while the second layer& regularization, while the second layer\\
  &consists of a linear prediction output unit.& is a softmax regression estimation layer.\\
\hline

6) & \multicolumn{1}{c}{Spectral parameters $T_\texttt{eff}$, log$~g$, [Fe/H], [C/H]} & \multicolumn{1}{|c|}{Category: carbon-normal metal-poor stars, CEMP, non-metal-poor}\\
\hline

\end{tabular}
}
\tablecomments{This model incorporates two branches, namely parameter estimation (described on the left side of the table) and classification (described on the right side of the table), in the last two rows (steps 5 and 6).}
\label{tab:ProposedScheme}

\end{table*}

\begin{figure}[htbp]
\centering
\setlength{\abovecaptionskip}{1pt}
\setlength{\belowcaptionskip}{6pt}

\setlength{\subfigbottomskip}{2pt}
\setlength{\subfigcapskip}{-9pt}

    \subfigure[$T_\texttt{eff}$]{\includegraphics[width=0.23\linewidth]{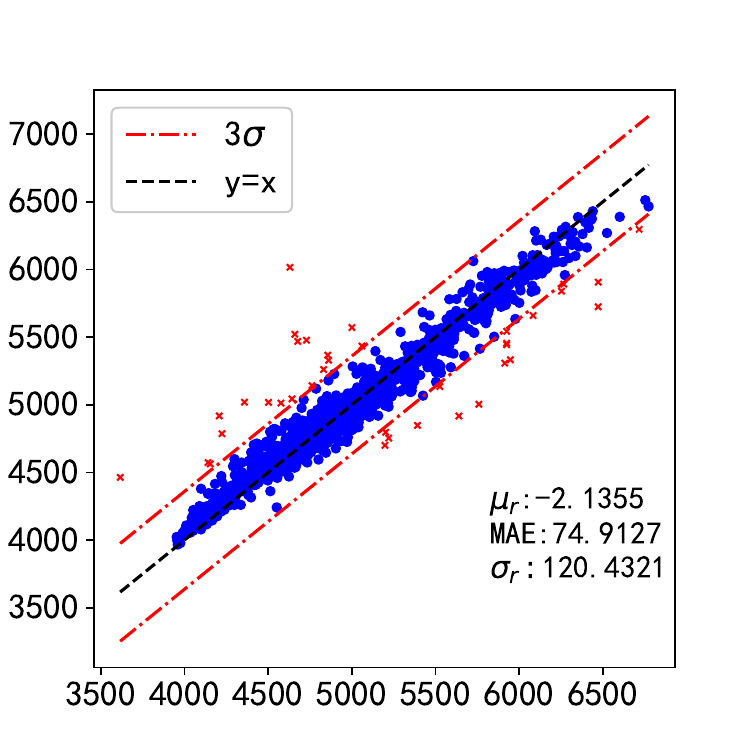}\label{fig:parameterEsitmiation4MP:Teff}}\hspace{0.01cm}
    \subfigure[log$~g$]{\includegraphics[width=0.23\linewidth]{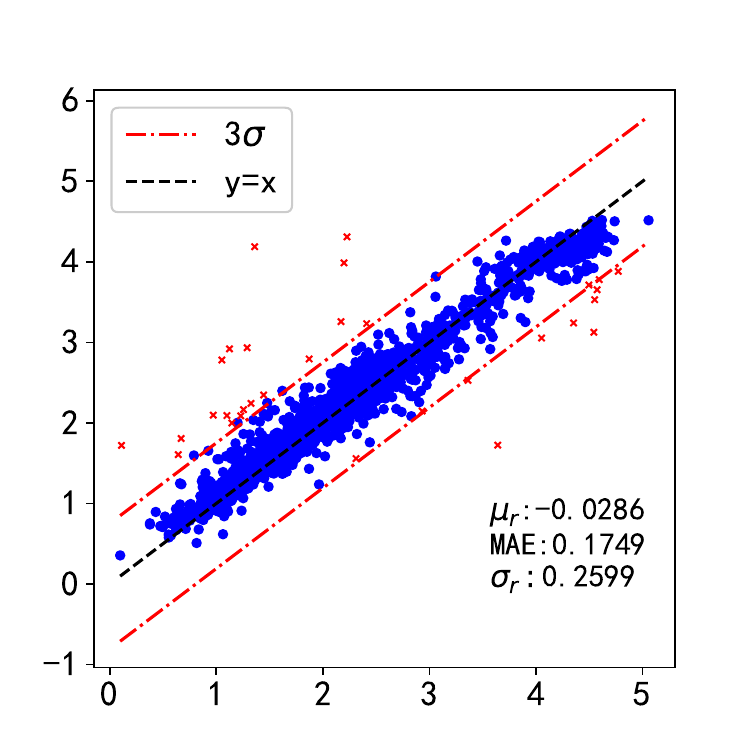}\label{fig:parameterEsitmiation4MP:logg}}
    \subfigure[{[}Fe/H{]}]{\includegraphics[width=0.23\linewidth]{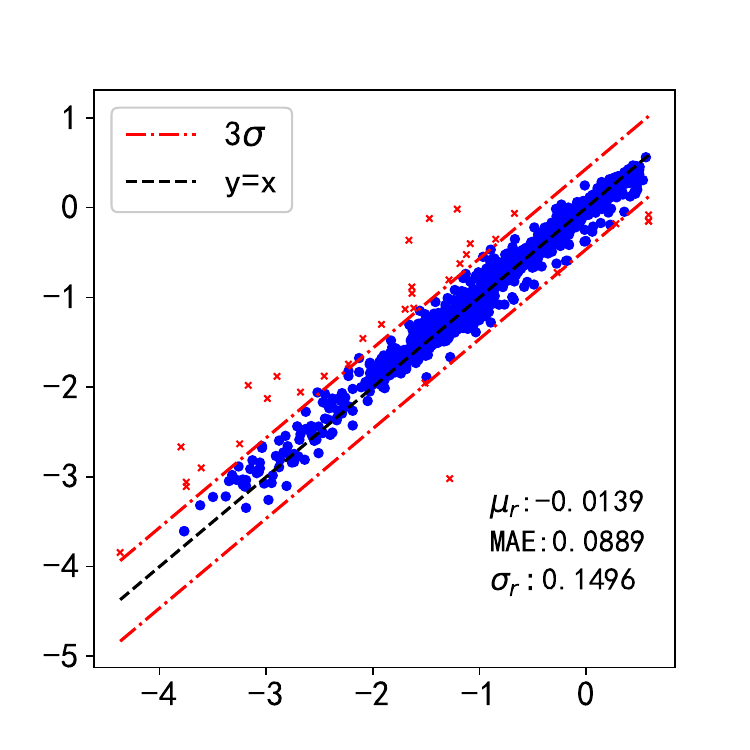}\label{fig:parameterEsitmiation4MP:FeH}}\hspace{0.01cm}
    \subfigure[{[}C/H{]}]{\includegraphics[width=0.23\linewidth]{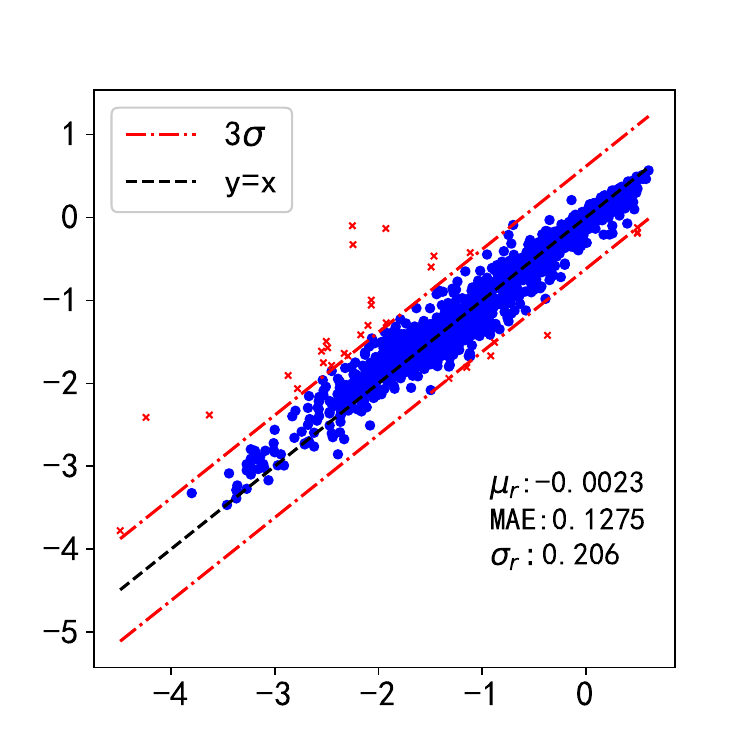}\label{fig:parameterEsitmiation4MP:CH}}\hspace{0.01cm}\hspace{0.01cm}
\caption{The performance of parameter estimation. The horizontal axis represents the reference values and the vertical axis represents the estimated values. These results are computed from the test set. The black dashed lines represent the reference line for theoretical consistency, while the red dashed lines represent the 3$\sigma$ reference lines. The $\mu_r$, $\sigma_r$ are the mean and standard deviation of the difference between the parameter estimation and its corresponing reference on test spectra. MAE is the mean absolute error/difference between the parameter estimation and its corresponing reference on test spectra.
}
\label{fig:parameterEsitmation4MP}
\end{figure}

\subsection{Model Evaluation}\label{test}

The learning process of the parameter estimation model proceeds as follows: first, the reference set $S^{stars}_{tr}$ (Section \ref{Sec:Workflow_Data_Preprocessing:Data4MP}) undergoes preprocessing (Section \ref{Sec:Workflow_Data_Preprocessing:Preprocessing}) to obtain the preprocessed training set $S^{stars}_{tr,pre}$. Subsequently, the preprocessed training set is fed into the MSFL-IECW Net to learn model parameters. The evaluation of the learning results on the test set $S^{stars}_{te}$ (Section \ref{Sec:Workflow_Data_Preprocessing:Data4MP}) are presented in Figure \ref{fig:parameterEsitmation4MP}.

For $T_\texttt{eff}$, $\log~g$, [Fe/H], and [C/H], only 1.89\%, 1.58\%, 1.74\%, and 1.58\% of the predictions deviate from their theoretical consistency over 3$\sigma$ respectively. This phenomenon indicates that this work is extremely consistent with the reference set. Due to the poor quality of a subset of the LAMOST spectra, it is unavoidable the existences of some parameter estimation with relatively evident deviations. This is the main reason for the outliers. Therefore, an additional high confidence catalog has been compiled in this work in section \ref{Sec:CEMPrecognition:results} in order to facilitate more accurate follow-up studies of CEMP stars. More related evaluations are conducted in section \ref{Sec:ParameterEvaluation}.

\section{
CEMP star search}\label{Sec:CEMPrecognition}
\subsection{
Search Results for CEMP Star Candidates}\label{Sec:CEMPrecognition:results}

Following the CEMP star search process in Section \ref{Sec:Workflow_Data_Preprocessing:Workflow}, we estimated stellar parameters $T_\texttt{eff}$, $\log~g$, [Fe/H], and [C/H] (Section \ref{Sec:ParaEst_MPsearching}). Subsequently, based on the [Fe/H] $\leq$ -1 criterion, we obtained some metal-poor star candidates. Then, according to the criteria in Equation (\ref{eq:CEMP_define}), we classified the metal-poor star candidates into CEMP star candidates and carbon-normal metal-poor star candidates.

\begin{figure}[htb]
    \centering
    \setlength{\abovecaptionskip}{1pt}
    \setlength{\belowcaptionskip}{6pt}

    \setlength{\subfigbottomskip}{10pt}
    \setlength{\subfigcapskip}{2pt}

    \subfigure[Distribution of CEMP candidates]{\includegraphics[width=0.46\textwidth]{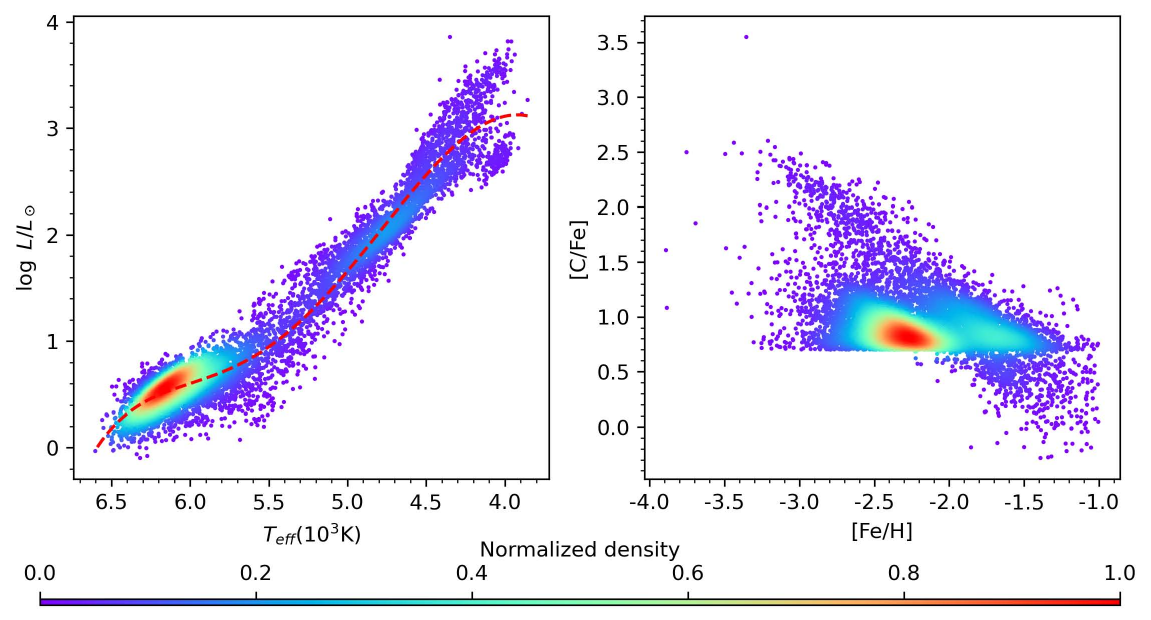}\label{fig:CEMP_distribution}}\hspace{0.01cm}
    \subfigure[Occurence rate of CEMP candidates]{\includegraphics[width=0.46\textwidth]{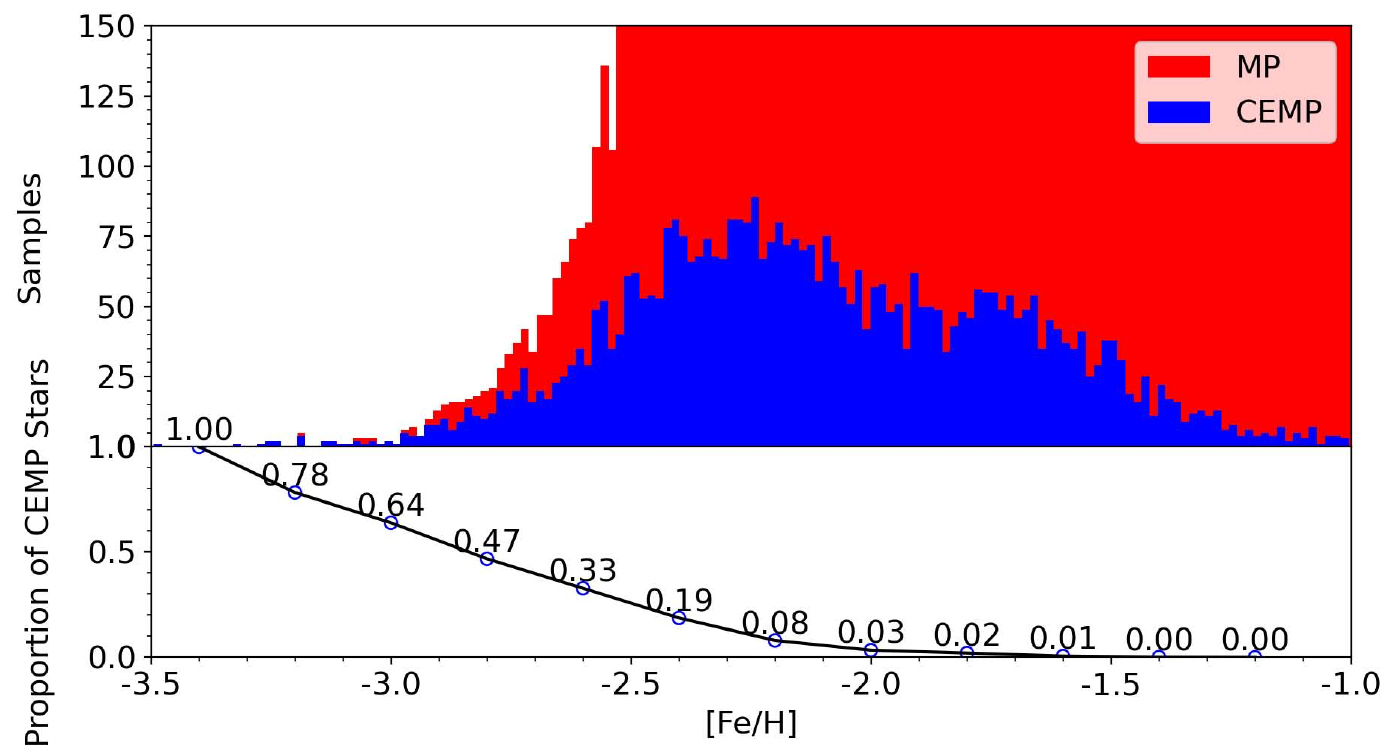}\label{fig:CEMP_proportion}}
    \caption{The distribution of 12,766 CEMP candidates and occurence rate of CEMP. In (b), proportions exceeding 150 are not displayed. [C/Fe] = [C/H] - [Fe/H].}
    \label{fig:CEMP_distribution_occurence_rate}
\end{figure}

By processing 8,651,552 low-resolution, {SNR$_g>$5} stellar spectra in LAMOST DR8, 12,766 CEMP star candidates were discovered. In these computed CEMP star candidates, 9,461 are  Very Metal-Poor (VMP) star candidates ([Fe/H] $<$ -2), and 164 are Extremely Metal-Poor (EMP) star candidates ([Fe/H] $<$ -3). The distribution of these CEMP star candidates in parameter space is presented in Figure \ref{fig:CEMP_distribution}.

According to the official documentation from LAMOST\footnote{\url{https://www.lamost.org/dr8/v2.0/doc/lr-data-production-description}}, some spectra are deemed unsuitable for deriving high-precision atmospheric physical parameters due to their quality issues. These spectral quality issues could potentially reduce the accuracy of the CEMP search.
Therefore, the CEMP star candidates corresponding to such spectra are annotated with low confidence, while the other 4,396 CEMP candidates are annotated with high confidence.

 The computed CEMP candidates are presented in a HR diagram (the left panel of Figure~\ref{fig:CEMP_distribution}). This subfigure illustrates the evolutionary sequence of the CEMP star candidates: transitioning from main sequence to subgiant, to red giant phases.
 While Equation~(\ref{eq:CEMP_define}) does not directly constrain the relationship between [C/Fe] and [Fe/H], the supplementary application of the CEMP classification formula to stars with [C/Fe] $<$ 0.7 is effective only for stars with [Fe/H] $>$ -2.5 as depicted in the right panel of Figure~\ref{fig:CEMP_distribution}. This phenomenon  is consistent with the observed distribution of the reference dataset.
The variation in carbon abundance is a result of the different evolutionary stages of stars. During evolution on the upper red giant branch, carbon in the stellar atmosphere can be converted into nitrogen through the carbon-nitrogen (CN) cycle, leading to its mixing into the stellar surface. This process results in carbon depletion, which consequently causes a decrease in the observable carbon abundance\citep{placco2014carbon}. As a result, some stars with lower carbon abundance in the later evolutionary stages have been reclassified as CEMP stars.

Figure~\ref{fig:CEMP_proportion} illustrates the relationship between the proportion of CEMP stars and [Fe/H]. As the metallicity decreases, the proportion of CEMP stars increases significantly. This trend is consistent with previous studies, such as \citet{placco2014carbon}.

\subsection{
Evaluation of the CEMP Star Searching Scheme }\label{Sec:CEMPrecognition:evaluation}

Based on the CEMP star search process in Section \ref{Sec:Workflow_Data_Preprocessing:Workflow}, along with the definition criteria for CEMP stars (Equation (\ref{eq:CEMP_define})), the effectiveness of CEMP star search depends on the  parameter estimation accuracy for $T_\texttt{eff}$, $\log~g$, [Fe/H], and [C/H]. Relevant evaluation results are presented in Figure \ref{fig:parameterEsitmation4MP}.

In this study, CEMP star spectra are regarded as positive samples, while carbon-normal metal-poor star spectra are considered as negative samples. We evaluated the effectiveness of the CEMP star search scheme using such metrics as accuracy, precision, recall, and F1 score.
Accuracy is used to measure the proportion of correctly identified CEMP and carbon-normal metal-poor stars. Recall is used to measure the proportion of correctly identified CEMP stars.
Precision is used to measure the proportion of correctly identified samples as CEMP among all samples classified as CEMP. In the context of this paper, the number of carbon-normal metal-poor stars is much higher than that of CEMP stars.
In such cases, even if accuracy is high, the precision or recall is possible very low. A low recall means that many CEMP stars fail to be detected. A low precision means that many carbon-normal metal-poor stars are mistakely recognized as CEMP star candidates. Therefore, we need a measure to comprehensively evaluate the recall and prcision. The F1 score provides a comprehensive measure of precision and recall; it is the harmonic mean of the two metrics.
Further discussion on the aforementioned metrics can be found in \citet{MNRAS:Zeng:2020}.

This work achieved an accuracy of 0.9554, a recall of 0.6970, a precision of 0.8519, and an F1 score of 0.7667 for CEMP search on the test set.
It can be observed that even though CEMP stars constitute a very small proportion of the total samples, this work can still identify approximately 69.7\% of the CEMP stars from a large number of carbon-normal metal-poor stars and non-metal-poor stars. Particularly, among the detected candidates, approximately 85.19\% of the sources are correctly identified as CEMP stars, which is highly valuable for subsequent high-resolution observation source selection.
In the future, as the reference data for confirmed CEMP stars increases, there is potential to further improve the identification effectiveness by expanding the training data and retraining the model presented in this paper.

\subsection{CEMP Searching Performance Evaluation Based on the CH G-band}

This work employs a data-driven approach to search for CEMP stars; we expect  the model's predictions to be consistent with astrophysical researches. \citet{LiHaiNing2018ApJS} demonstrates that the G1 band line index is highly sensitive to carbon enhancement. Therefore, it helps evaluate the accuracy of CEMP star candidate search to analyze the flux in the G1 band. \citet{Placco2011AJ} proposed the EGP index to measure the CH G-band. This measure helps identify CEMP star candidates from low-resolution spectra while avoiding contamination from the strong $H_\gamma$ line. The calculation method for the EGP is provided in Equation~(\ref{eq:EGP}).

\begin{equation}
\label{eq:EGP}
\begin{aligned}
\mathrm{EGP}=-2.5 \log \left(\frac{\int_{4200}^{4400} I_{\lambda} d \lambda}{\int_{4425}^{4520} I_{\lambda} d \lambda}\right)
\end{aligned}
\end{equation}

\begin{figure}[htb]
    \centering
    \setlength{\abovecaptionskip}{1pt}
    \setlength{\belowcaptionskip}{-2pt}

    \setlength{\subfigbottomskip}{1pt}
    \setlength{\subfigcapskip}{-6pt}

    \includegraphics[width=0.8\linewidth]{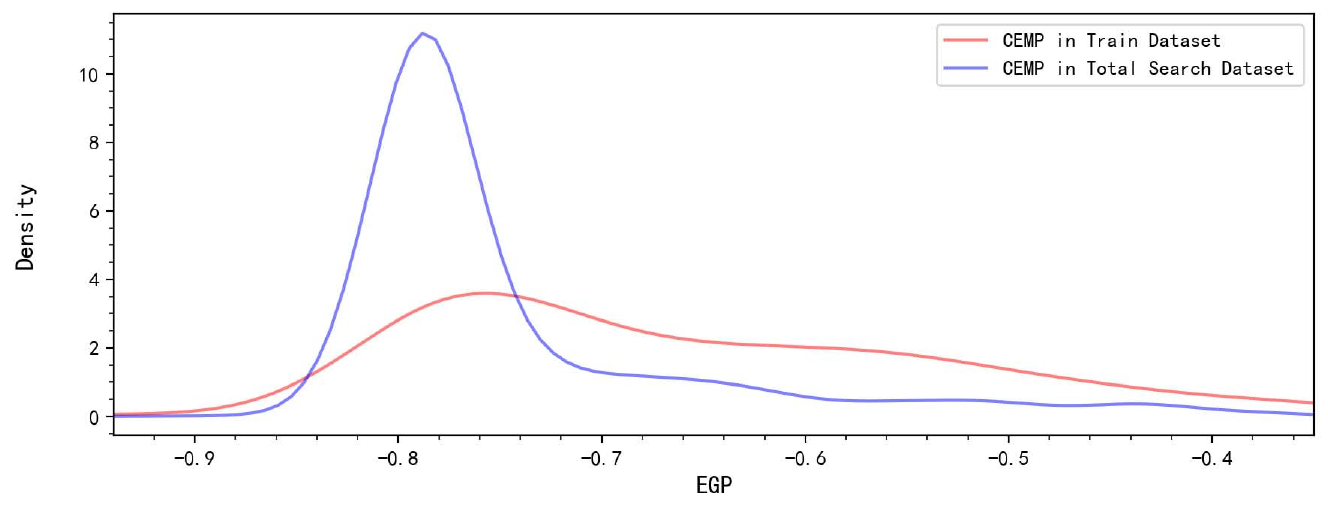}\hspace{0.01cm}
    \caption{Distribution of EGP indices for CEMP stars, with blue for all CEMP stars searched and red for CEMP stars in the training set.}

    \label{fig:EGP_rate}
\end{figure}

Experimental results in Figure \ref{fig:EGP_rate} shows the distribution of the discovered CEMP star candidates in the EGP index space, where the blue curve represents the distribution of the CEMP stars in the training set and the red curve represents the distribution of the CEMP star candidates discovered in this work. The CEMP stars in the training set were carefully selected from several literatures, and verified by medium-resolution or high-resolution spectroscopic observations.
Both CEMP star datasets have a peak of sample occurence probability on $ -0.85 \leq \texttt{EGP} \leq-0.7$, and their occurence probability of CEMP star gradually decrease with the distance increases from the peak in the cases of $ \texttt{EGP}> -0.7$ and $ \texttt{EGP} <-0.85$. Therefore, their EGP occurence ranges and distribution characteristics are consistent to some degree. This consistency indicates the reliability of the CEMP star candidate searched in this work.
At the same time, there is a very strong selection effect in the medium-resolution or high-resolution observation verification stage. Therefore, there exist some difference in their distributions.

\begin{figure}[htb]
    \centering
    \setlength{\abovecaptionskip}{1pt}
    \setlength{\belowcaptionskip}{-2pt}

    \setlength{\subfigbottomskip}{1pt}
    \setlength{\subfigcapskip}{-6pt}

    \subfigure[On samples from APOGEE DR17]{\includegraphics[width=0.32\linewidth]{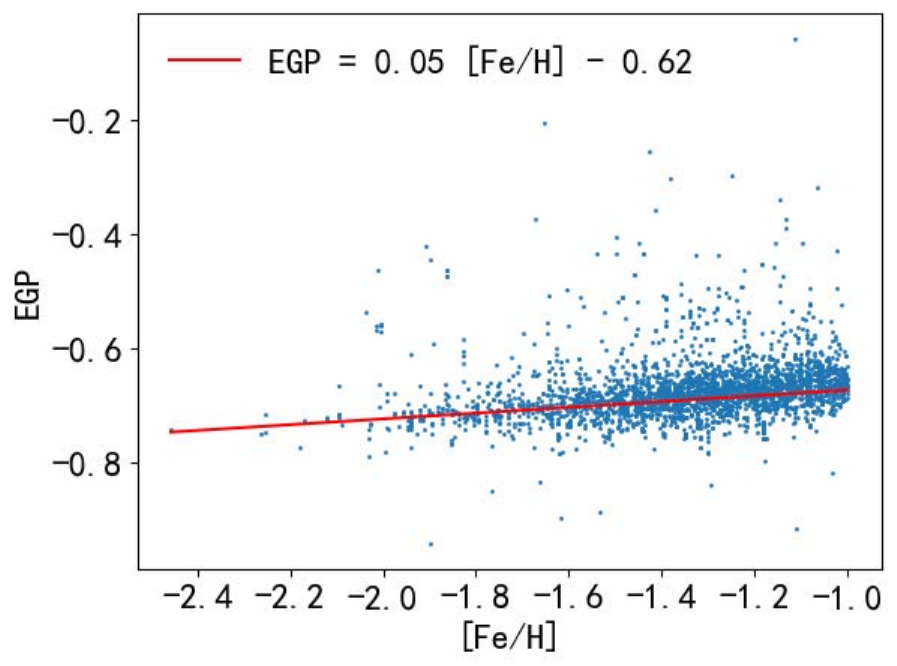}}\label{fig:Dependencies_EGP_FeH:case1:Apogee}\hspace{0.01cm}
    \subfigure[On samples from LAMOST-Subaru.]{\includegraphics[width=0.32\linewidth]{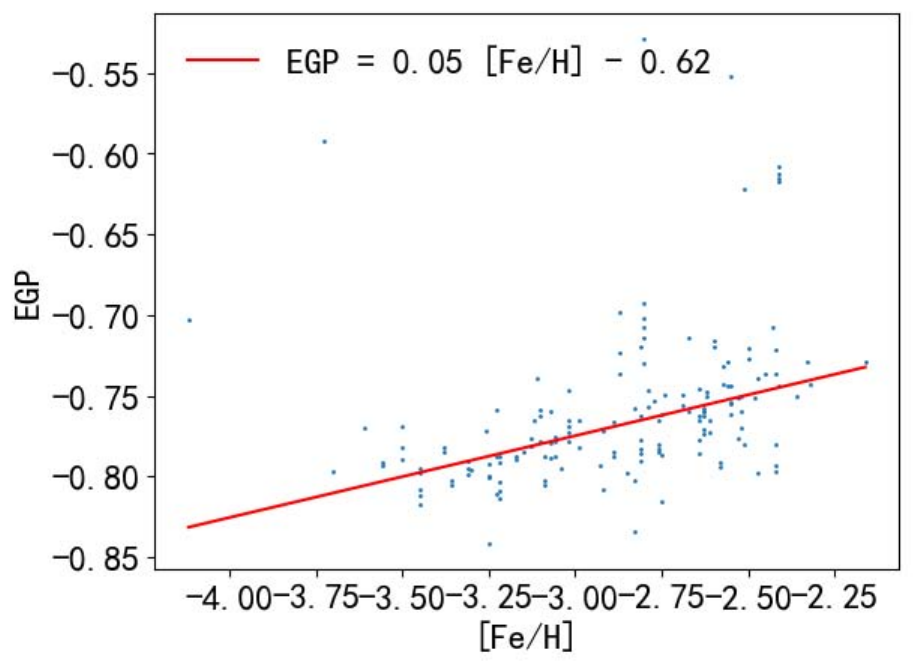}}\label{fig:Dependencies_EGP_FeH:case1:Subaru}\hspace{0.01cm}
    \subfigure[On samples of this work.]{\includegraphics[width=0.32\linewidth]{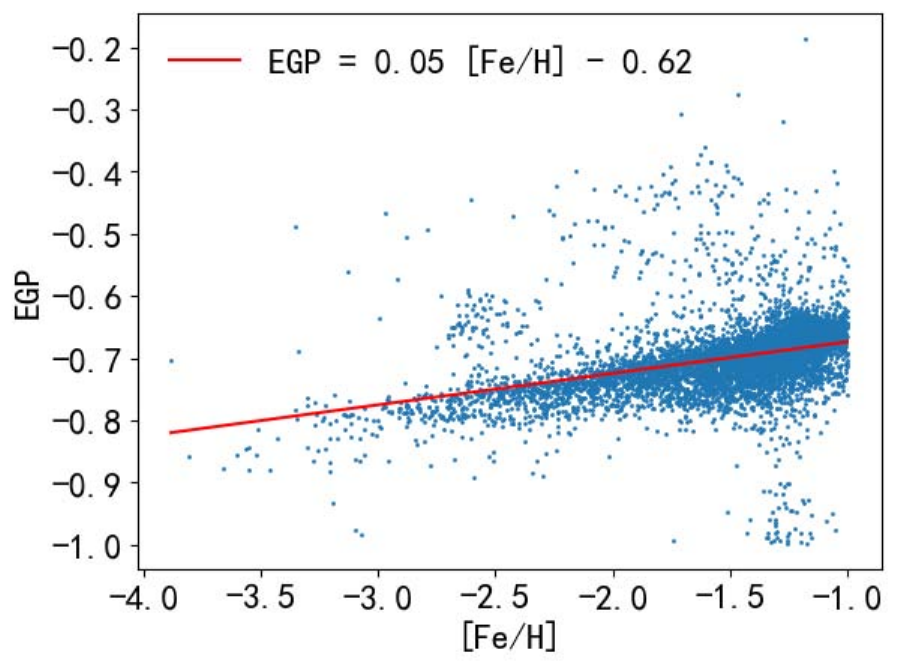}}\label{fig:Dependencies_EGP_FeH:case1:thisWork}

    \subfigure[On samples from APOGEE DR17]{\includegraphics[width=0.32\linewidth]{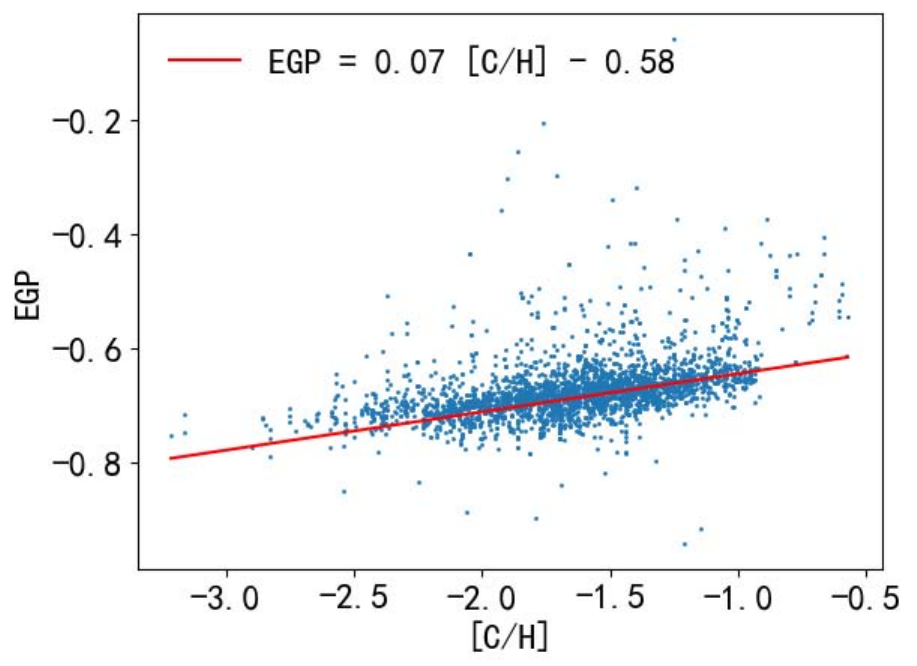}}\label{fig:Dependencies_EGP_CH:case1:Apogee}\hspace{0.01cm}
    \subfigure[On samples from LAMOST-Subaru]{\includegraphics[width=0.32\linewidth]{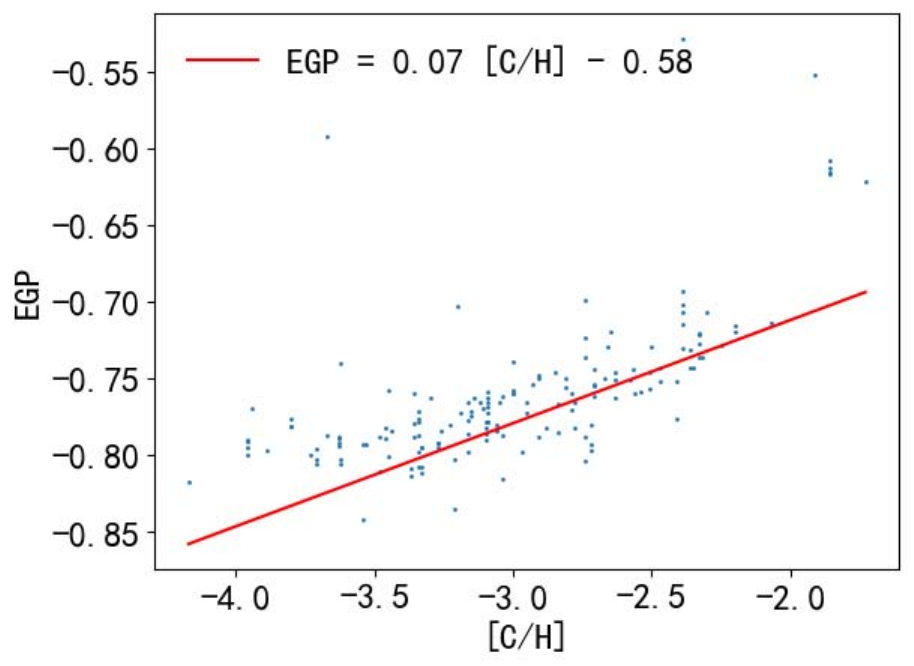}}\label{fig:Dependencies_EGP_CH:case1:Subaru}\hspace{0.01cm}
    \subfigure[On samples of this work.]{\includegraphics[width=0.32\linewidth]{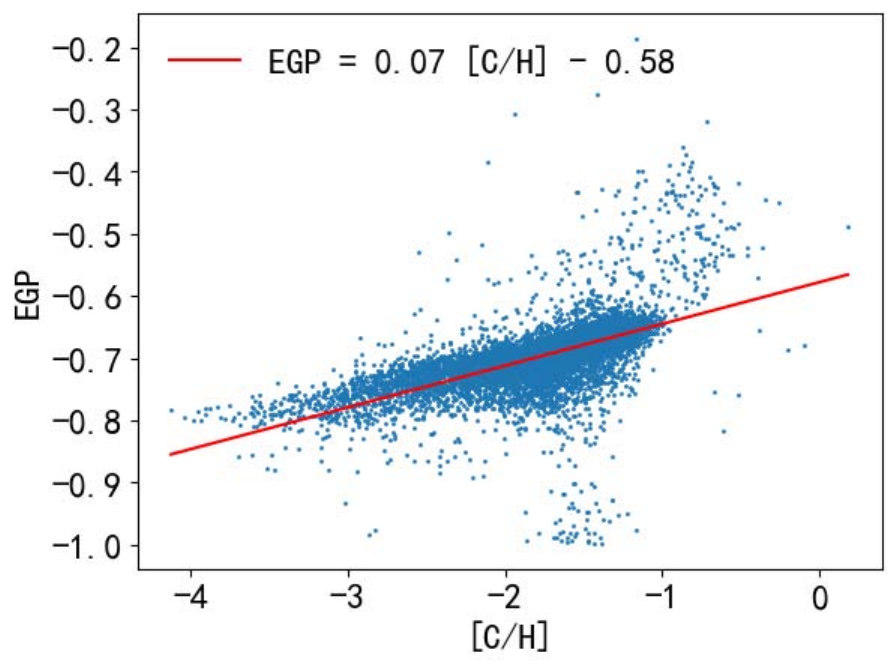}}\label{fig:Dependencies_EGP_CH:case1:thisWork}

    \caption{Dependences of EGP index on [Fe/H] and [C/H] on the cases of $4300K\leq T_\texttt{eff}\leq$ 5000K \& $\log~g \leq$ 2.5. The lines in (a), (b),(c) are estimated using the samples in (c), similarly for the lines in (d), (e) and (f).
     }
    \label{fig:Dependencies_EGP_FeHCH}
\end{figure}

Furthermore, the dependencies of EGP index on [Fe/H] and [C/H] are investigated respectively in Fig. \ref{fig:Dependencies_EGP_FeHCH}.
It is shown that the EGP index linearly depends on [C/H], and [Fe/H] excellently in cases of $4300K\leq T_\texttt{eff} \leq$5000K \& $\log~g \leq $2.5, and $T_\texttt{eff} >5000K$ \& $\log~g\leq 4.5$.

\section{Comparing the Parameter Estimation Results with GALAH}\label{Sec:ParameterEvaluation}

The Third Data Release of the Galactic Archaeology with HERMES (GALAH DR3) \citep{buder2021galah+} contains high-quality spectroscopic data for nearly 600,000 stars, covering a wide range of stellar types from dwarfs to giants, as well as a broad range of metal abundances from metal-poor to metal-rich stars. These spectra have a resolution of up to 28,000 and allow for precise measurements of the abundances of up to 30 elements. GALAH utilizes the spectral grid from AMBRE for global fitting to obtain initial parameters, and refines parameter estimates by linearly combining parameters from the ten closest model spectra.

\begin{figure}[htbp]
    \centering
    \setlength{\abovecaptionskip}{1pt}
    \setlength{\belowcaptionskip}{6pt}
    \setlength{\subfigbottomskip}{2pt}
    \setlength{\subfigcapskip}{-5pt}
    \subfigure[$T_\texttt{eff}$]{\includegraphics[width=0.23\linewidth]{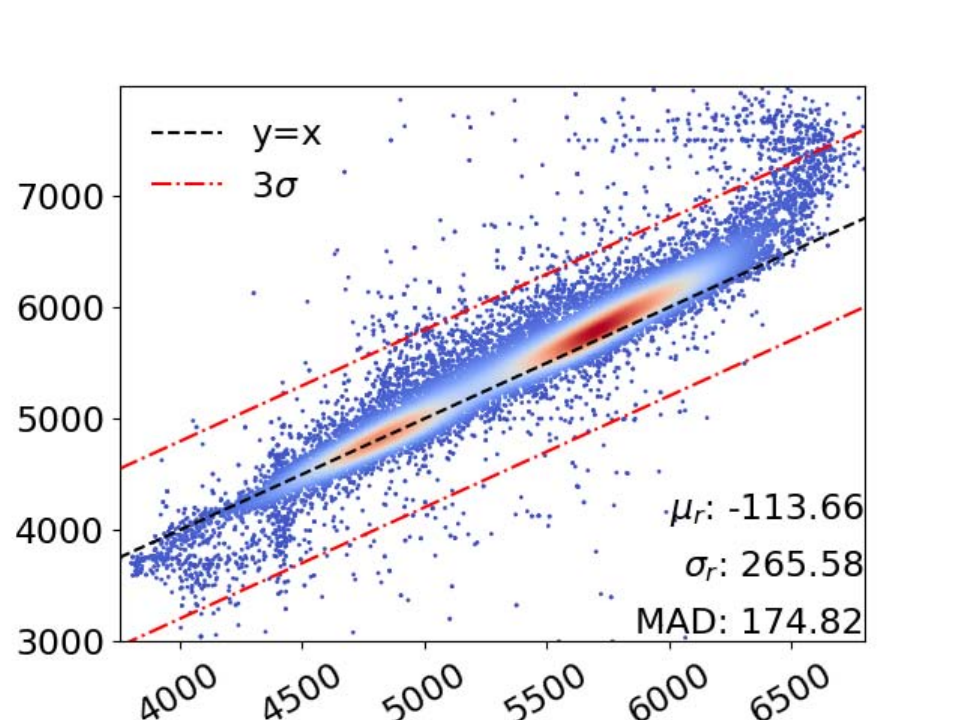}\label{fig:galah_compare:Teff}}\hspace{0.01cm}
    \subfigure[log$~g$]{\includegraphics[width=0.23\linewidth]{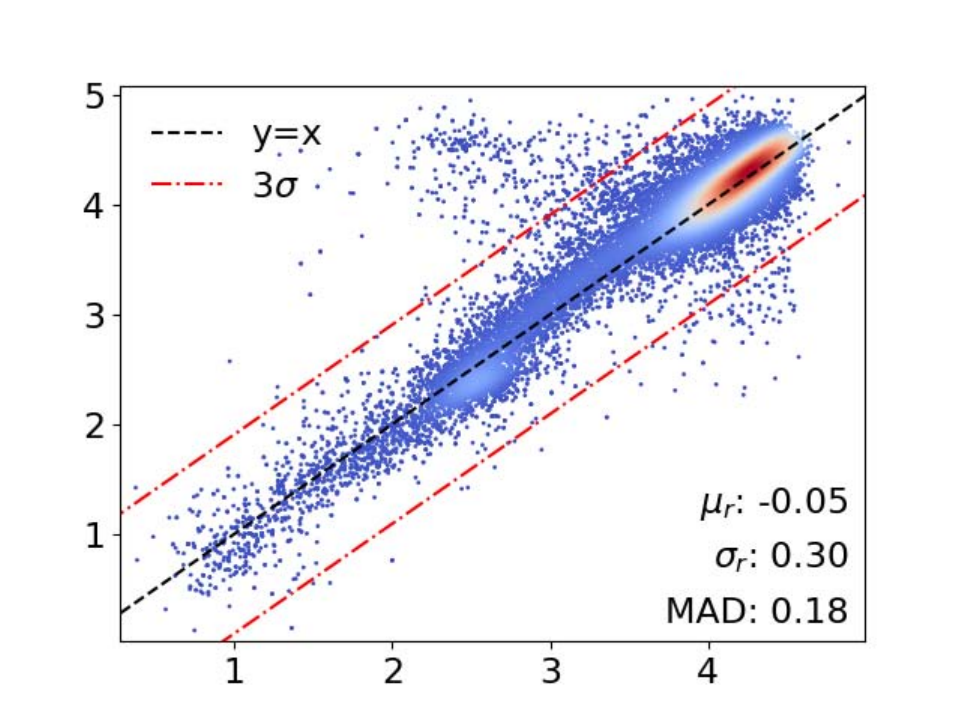}\label{fig:galah_compare:logg}}\hspace{0.01cm}
    \subfigure[{[}Fe/H{]}]{\includegraphics[width=0.23\linewidth]{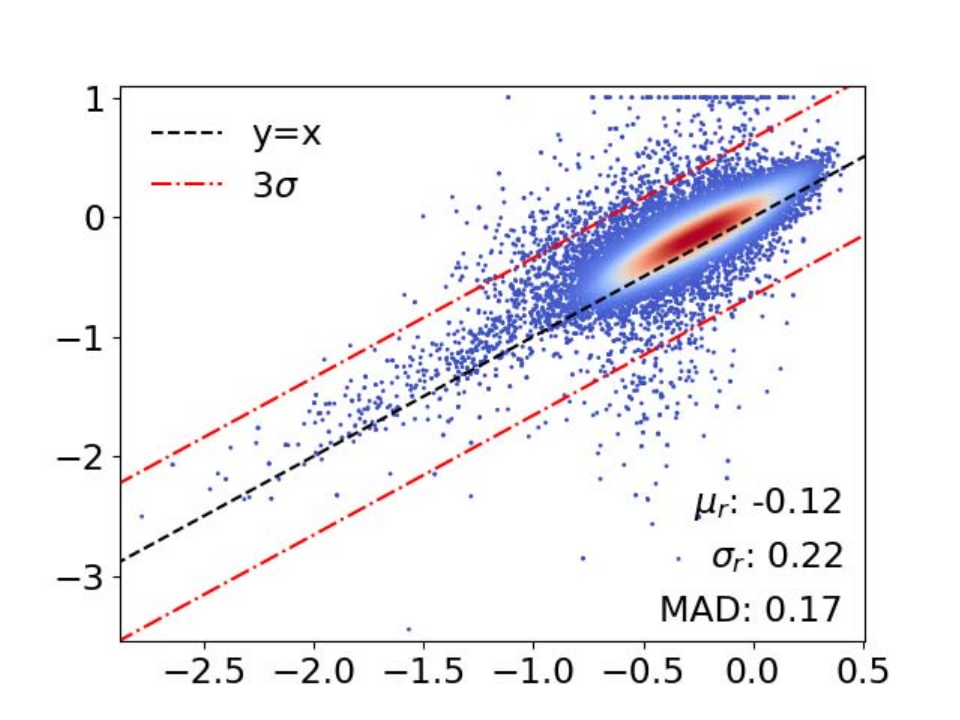}\label{fig:galah_compare:FeH}}\hspace{0.01cm}
    \subfigure[{[}C/H{]}]{\includegraphics[width=0.23\linewidth]{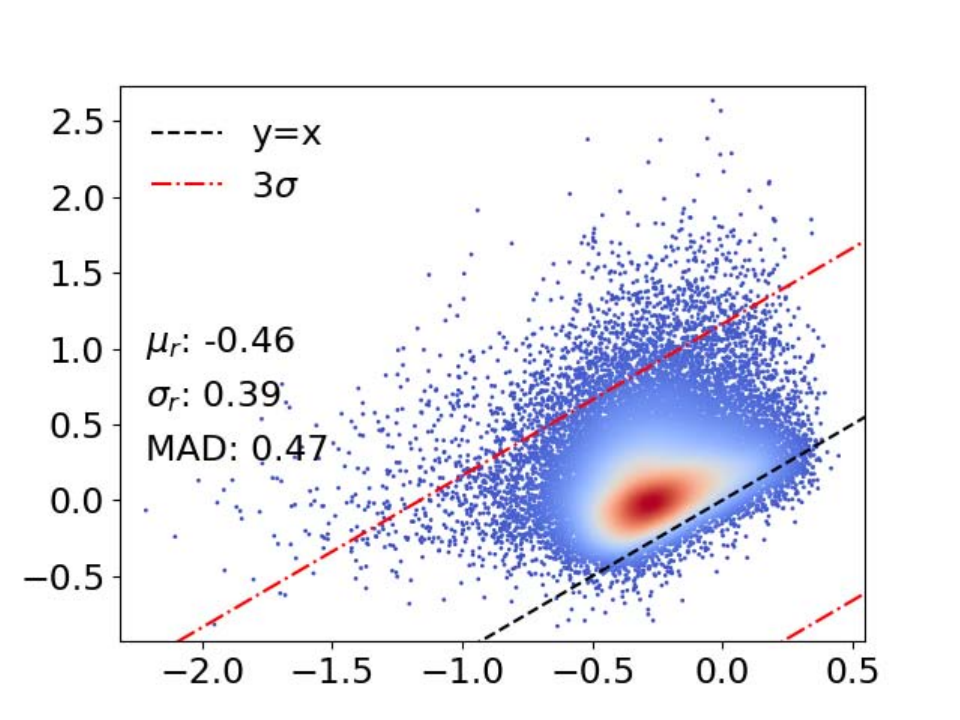}\label{fig:galah_compare:CH}}
    \caption{The comparisons between this work and GALAH. The x-axis represents the parameter estimates from this work, while the y-axis represents the GALAH estimates. The color intensity corresponds to the density of samples, with brighter colors indicating higher sample densities. The black dashed line represents the line of theoretical consistency, while the red dashed line represents the 3$\sigma$ reference line.The $\mu_r$, $\sigma_r$ are the mean and standard deviation of the difference between two catalogs. MAD is the mean absolute difference between two catalogs.The horizontal coordinates of the graphs indicate the EGP values, the vertical coordinates indicate the density of the distribution, and the area of each bar graph sums to one.}

    \label{fig:galah_compare}
\end{figure}

According to the discriminant criterion (\ref{eq:CEMP_define}), the effectiveness of CEMP star searches depends on the estimation of four parameters: $T_\texttt{eff}$, $\log~g$, [Fe/H], and [C/H]. Consistency with parameter estimates from high-resolution spectroscopic catalogs can be a good indication of the credibility of this work. In Section \ref{test}, our work demonstrates a high level of consistency with reference catalogs, such as APOGEE DR17 \citep{ApJS:Abdurro:2022}, SAGA \citep{PASJ:Suda:2017}, LAMOST-Subaru \citep{Li_2022}. Furthermore, in Section \ref{Sec:CEMPrecognition:evaluation}, we compare the parameter estimation results with GALAH \citep{buder2021galah+} to indirectly evaluate the reliability of our search results. The comparison between GALAH and our work is presented in Figure \ref{fig:galah_compare}.

For the estimation of effective temperature, there is a divergence branch on the GALAH parameter range of [6773K,8000K] (Figure~\ref{fig:galah_compare:Teff}).
This discrepancy arises from the fact that our training samples predominantly concentrates within the temperature range of [3615K, 6773K]; consequently, our model may exhibit a certain degree of underestimation for samples out of this range.
Significant temperature discrepancies between the two studies occur on only about 0.4\% of the samples. These discrepany samples primarily concentrate in the high-temperature region. In the high-temperature discrepancy region, no CEMP stars are found, this phenomenon is consistent with expectations. Carbon signals notably become weak as temperature increases. This case is consistent with the findings of \citet{witten2022information}, indicates that [C/H] precision can only reach 0.5 in the case of $T_\texttt{eff}$ $<$ 6000 K. It is challenging to estimate carbon enhancement for stars with excessively high temperatures. Therefore, the discrepancy between this work and GALAH in the high-temperature region does not affect the result for CEMP star searches.

For the estimation of $\log~g$, this work exhibits a relatively high consistency with GALAH, although approximately 0.4\% of stars are underestimated near $\log~g$ $\approx 5.0$. No CEMP stars were found at the divergence area. This result is  predictable. This predictability is due to the fact that the largest $\log~g$ of CEMP stars in the reference set of this work is about 4. Therefore, in case of a CEMP sample with a larger $\log~g$, this work may suffer from underestimation.

In terms of [Fe/H] estimation, the two works are in good agreement overall and especially for [Fe/H] $<$ -2. This excellent consistency indicates a high credibility of the selected VMP candidates. CEMP stars are predominantly found among VMP stars. Therefore, the high accuracy of VMP stars enables us to confidently identify CEMP stars in this work.

For the estimation of [C/H], this work exhibits significant systematic biases compared to GALAH, similar biases are also observed between APOGEE DR17 and GALAH. This may stem from the inconsistencies in the standards used for [C/H] between the two studies.

In summary, this work demonstrates a high level of consistency with GALAH on $T_\texttt {eff}$ for $[3615.00, 6772.54]$ K, $\log~g$ for $[-0.15, 5.06]$, [Fe/H] for $[-4.38, 0.59]$ and [C/H] for $[-4.50, 0.90]$. Particularly, the accuracy of parameter estimation for VMP stars ensures the credibility of the CEMP stars identified in our study.

\section{Conclusion}\label{Sec:Conclusion}
This study addresses the search for Carbon-Enhanced Metal-Poor (CEMP) stars based on low-resolution spectra and proposes a machine learning solution based on deep learning/neural networks. Initially, the method estimates stellar parameters  $T_\texttt{eff}$, $\log~g$, [Fe/H], and [C/H] from observed spectra. The CEMP star candidates are identified based on the estimated parameters.
Using this approach, we discovered 819,671 metal-poor star candidates and 12,766 CEMP star candidates from the LAMOST DR8 low-resolution stellar spectrum database. Among these CEMP star candidates, there are 9,461 VMP star candidates and 164 EMP star candidates. For ease of reference and use, we provide estimates of such parameters as $T_\texttt{eff}$, $\log~g$, [Fe/H], and [C/H] for the identified targets. The computed catalog is available at \url{https://nadc.china-vo.org/res/r101500/}.

\vspace{0.4cm}
\textbf{Acknowledgments~~}
This work is supported by the National Key R\&D Program of China (Grant No. 2024YFA1611903),  the National Natural Science Foundation of China (Grants Nos. 12373108 and 12222305). Guoshoujing Telescope (the Large Sky Area Multi-Object Fiber Spectroscopic Telescope, LAMOST) is a National Major Scientific Project built by the Chinese Academy of Sciences.

\bibliography{sample631}{}
\bibliographystyle{aasjournal}



\end{document}